\documentclass{amsart}
\usepackage{amssymb,amssymb}
\usepackage{mathrsfs}

\theoremstyle{plain}
\newtheorem{theorem}{Theorem}[section]
\newtheorem{lemma}[theorem]{Lemma}
\newtheorem{proposition}[theorem]{Proposition}

\theoremstyle{remark}

\newtheorem{example}[theorem]{Example}

\theoremstyle{plain}
\newtheorem{corollary}[theorem]{Corollary}
\numberwithin{equation}{section}

\renewcommand{\H}{{\mathscr H}}
\renewcommand{\Re}{{\rm Re}\,}
\renewcommand{\Im}{{\rm Im}\,}

\newcommand{\e}{\varepsilon}

\newcommand{\g}{\gamma}
\renewcommand{\l}{z}

\newcommand{\n}{\Vert}
\newcommand{\lb}{\langle}
\newcommand{\rb}{\rangle}
\newcommand{\s}{^{\ast}}
\newcommand{\limn}{\lim_{n\to\infty}}
\newcommand{\G}{\Gamma}
\renewcommand{\odot}{\hbox{\tiny\textcircled{s}}}
\newcommand{\Gs}{\G^{\odot}}
\newcommand{\Hs}{{\G}^{\odot}(H)}
\newcommand{\embed}{\hookrightarrow}

\newcommand{\calL}{{\mathscr L}}
  
\renewcommand{\sp}{\sigma}
\newcommand{\N}{{\mathbb N}}
\newcommand{\R}{{\mathbb R}}
\newcommand{\C}{{\mathbb C}}
\newcommand{\Z}{{\mathbb Z}}

\begin{document}

\title[$L^p$-Spectrum of Ornstein-Uhlenbeck operators]
{Second Quantization and the $L^p$-spectrum of 
nonsymmetric Ornstein-Uhlenbeck operators}

\author{J.M.A.M. van Neerven}
\address{Delft Institute of Applied Mathematics\\
Technical University of Delft \\ P.O. Box 5031\\ 2600 GA Delft\\The Netherlands}
\email{J.vanNeerven@math.tudelft.nl}
 
\begin{abstract} 
The spectra of the second quantization and the 
symmetric second quantization
of a strict Hilbert space contraction are computed explicitly and
shown to coincide.
As an application, we compute the spectrum of the
nonsymmetric Ornstein-Uhlenbeck operator $L$
associated with the infinite-dimensional
Lange\-vin equation
$$ dU(t) = AU(t)\,dt + dW(t)$$
where $A$ is the generator of a strongly continuous semigroup
on a Banach space $E$ and $W$ is a cylindrical Wiener 
process in $E$. 
Assuming the existence of an invariant measure $\mu$ for $L$,
under suitable assumptions on $A$ we show that the spectrum of $L$ 
in the space $L^p(E,\mu)$ ($1<p<\infty$)
is given by
$$\sp(L) = \overline{\Bigl\{ \sum_{j=1}^n k_j z_j: \ k_j\in\N, \ z_j
\in\sp(A_\mu);\ j=1,\dots,n; \ n\ge 1\Bigr\}},
$$
where $A_\mu$ is the generator of a Hilbert space contraction
semigroup canonically associated with $A$ and $\mu$.
We prove that the assumptions on $A$ 
are always satisfied in the strong Feller case and 
in the finite-dimensional case. In the latter case we recover the 
recent Metafune-Pallara-Priola formula for $\sp(L)$.
\end{abstract}

\keywords{Symmetric Fock space, 
second quantization, nonsymmetric Ornstein-Uhlenbeck operators, 
$L^p$-spectrum, invariant measure, reproducing kernel Hilbert spaces,
strong Feller property}
\subjclass[2000]{Primary: 35R15, 81S10; Secondary: 35P05, 47D03, 47D07,
60H15}
 
\thanks{This work was supported by ARC Discovery Grant DP0346406,
a `VIDI subsidie' in the `Vernieuwingsimpuls' programme of
the Netherlands Organization for Scientific Research (NWO) 
and by the Research Training Network HPRN-CT-2002-00281}

\maketitle

\section{Introduction}

There has been a considerable recent interest 
in the spectral theory of second order elliptic operators $L$ of the form
\begin{equation}\label{0}
L\phi(x) =\tfrac12 \hbox{Tr}\, (Q(x)D^2 \phi(x))+ \lb A(x),
D\phi(x)\rb\qquad (x\in E)
\end{equation}
with unbounded drift term $A$
on a finite or infinite dimensional space $E$; see for example
\cite{EckHai, Me, MP, MPP, MRS}.
In many situations $L$ admits a unique invariant measure $\mu$, in which case
it is natural to
consider the realization of $L$ in the space $L^p(E,\mu)$; see \cite{ABR, BRS,
CG1, DZ2}
and the references cited there.
Even in space dimension one this class of operators
is not completely understood at present.

In this paper we consider the case of Ornstein-Uhlenbeck operators,
i.e., the special case of \eqref{0} where $Q(x)=Q$ (the `diffusion') is a
fixed positive symmetric operator from $E\s$ into
$E$
and $A(x)=Ax$ with $A$ (the `drift') an
infinitesimal generator of a strongly continuous semigroup of operators on
$E$. The state space $E$ is allowed to be an arbitrary real Banach space,
the operator $Q$ is not assumed to have finite trace and
the operator $A$ may be unbounded.
We do not assume that $L$ is symmetric. Nonsymmetric Ornstein-Uhlenbeck 
operators arise naturally as the infinitesimal generators
of transition semigroups associated with stochastic partial differential 
equations and have been applied for example to
optimal control problems and to interest rate models; see 
\cite{DZ, DZ2, DZ3, GM} and the references cited therein. 
In finite dimensions,
nonsymmetric Ornstein-Uhlenbeck 
operators have been recently applied 
in the area of nonequilibrium statistical physics \cite{BLL}. 

Assuming the existence of an invariant measure $\mu$ for $L$,
our aim is to determine the spectrum of $L$
in $L^p(E,\mu)$ for $p\in (1,\infty)$.
Let us recall that in $E=\R^d$, the `classical' Ornstein-Uhlenbeck
operator with $Q=I$ and $A = -I$,
\[L\phi (x)=\tfrac 12 \Delta\phi(x) -\lb x, \nabla\phi(x)\rb\qquad (x\in\mathbb R^d)\]
which arises in quantum field theory as the boson number operator,
has a unique Gaussian invariant measure $\mu$ and the spectrum
of $L$ in $L^p(\R^d,\mu)$ is given by
$$
\sigma (L)=\left\{-n: \ n\in\N\right\}
$$
where $\N = \{0,1,2,\dots\}$. This formula
is an easy consequence of the description of $L$ as a second quantized
operator \cite{Ja, Pa, Si} 
and can be extended without much difficulty to the case where
$E$ is a Hilbert space
and $A$ is selfadjoint with compact resolvent.
The nonsymmetric case is considerably more difficult, however,
even for $E=\mathbb R^d$.
Under suitable nondegeneracy assumptions on $A$ and $Q$ it was shown
by Metafune, Pallara and Priola \cite{MPP} that the spectrum of 
$L$ in $L^p(\R^d,\mu)$ is given by
\begin{equation}\sigma (L)=\Bigl\{\sum_{i=1}^n k_j z_j: \ k_j\in\N, \
z_j\in\sigma(A);\ j=1,\dots,n; \ n\ge 1  \Bigr\}.\label{02}
\end{equation}
In particular the spectrum is independent of $p\in (1,\infty)$.
On the other hand it was shown by Metafune \cite{Me} that the spectrum of
$L$ in $L^p(\R^d)$ is $p$-dependent. This contrasts well-known results 
on spectral $p$-independence in $L^p(\R^d)$ of second-order elliptic 
operators under various different assumptions; see for example \cite{Dav, HeVo} 
and the references cited therein.

The proof of \eqref{02}  in \cite{MPP} 
depends on a careful analysis of the smoothing effects of the
transition semigroup $P=\{P(t)\}_{t\ge 0}$ generated by $L$.
In this paper we will give a completely different proof of an infinite dimensional version of
\eqref{02}  which instead exploits the fact that the transition
semigroup can be represented as the
symmetric second quantization of the adjoint of an appropriate
nonsymmetric contraction semigroup $S_\mu=\{S_\mu(t)\}_{t\ge 0}$
acting on the reproducing kernel Hilbert space
of the invariant measure $\mu$.
The crucial step in this approach is to obtain a formula for the spectrum of the
symmetric second quantization of Hilbert space contractions $T$.
For strict contractions $T$ this problem is
solved completely. The main difficulty consists of
showing that the spectra
of the $n$-fold tensor product and the symmetric $n$-fold tensor
product of $T$ coincide and are given by
$$ \sp(T^{\odot n}) = \sp(T^{\otimes n}) = \Bigl\{\prod_{j=1}^n z_j: \ z_j\in
\sp(T); \ j=1,\dots,n\Bigr\},$$
the second equality being a classical result due to Brown and Pearcy \cite{BP}.
This easily implies equality of the spectra of the second quantization and the
symmetric second quantization of $T$:
$$\sp(\Gs(T)) = \sp(\G(T))  = \{1\}\cup\overline{
\bigcup_{n\ge 1}\Bigl\{\prod_{j=1}^n z_j: \ z_j\in
\sp(T); \ j=1,\dots,n\Bigr\}}.$$
As a result we are able to compute the spectra of the operators $P(t)$ in
$L^2(E,\mu)$ under the assumption that $S_\mu$ is a semigroup of strict
contractions. By combining these arguments with  standard
hypercontractivity results we obtain
the spectra of $P(t)$ in  $L^p(E,\mu)$  for all $p\in (1,\infty)$.
The spectrum of $L$ in  $L^p(E,\mu)$ is then obtained via spectral mapping
techniques. For this step we require that, in addition to being
strictly contractive, $S_\mu$ is also eventually norm continuous.
Our main result asserts that under these assumptions
(which are shown to be automatically satisfied in two important cases:
the strong Feller case and the finite dimensional case), we have
$$
\sigma (L)=\overline{\Bigl\{\sum_{i=1}^n k_j z_j: \ k_j\in\N, \
z_j\in\sigma(A_\mu); \ j=1,\dots,n; \ n\ge 1  \Bigr\}}.
$$
Here, $A_\mu$ denotes the generator of the semigroup
$S_\mu$. If $S_\mu$
is compact (which is the case in the strong Feller case
and in finite dimensions), no closure needs to be taken and we obtain
\begin{equation}\sigma (L)=\Bigl\{\sum_{i=1}^n k_j z_j: \ k_j\in\N, \ 
z_j\in\sigma(A_\mu); \ j=1,\dots,n; \ n\ge 1  \Bigr\}.\label{04}
\end{equation}
In finite dimensions, under a nondegeneracy assumption we 
have $\sp(A_\mu) = \sp(A)$ and 
\eqref{04} reduces to the Metafune-Pallara-Priola formula \eqref{02}.

\medskip
{\em Acknowledgments} -
This work was done while the author stayed
at the University of New South Wales. He thanks his colleagues at the School 
of Mathematics, especially Ben Goldys and Ian Doust, 
for their kind hospitality. This paper owes its existence to the numerous
inspiring discussions with Ben Goldys who also pointed out the crucial references
\cite{He} and \cite{RS}. 
The author thanks Marco Furhman and Silvania
Pereira for helpful comments.

\section{Preliminaries}\label{sec2}

In this section we collect some well known results on
spectral theory and reproducing kernel 
Hilbert spaces. For more detailed information we refer to \cite{Au, Bo}.

\subsection{Spectral theory} 
Let $X$ be a real or complex Banach space. 
The spectrum of a bounded or unbounded linear operator $T$
on $X$ will be denoted by
$\sp(T)$.  When $X$ is a real Banach space, the spectrum of $T$ is 
defined as the spectrum of its complexification.
The spectral radius of a bounded
operator $T$ is denoted by $r(T)$. 
The point spectrum, approximate point spectrum, and residual 
point spectrum of $T$ will be denoted by $\sigma_{\rm p}(T)$, 
$\sp_{\rm a}(T)$, and $\sp_{\rm r}(T)$ respectively; the latter is
defined as the set of all $\l\in \sp(T)$ for which
the range of  $\l-T$ is a proper, closed subspace of $X$.
Recall that $\partial \sp(T)\subseteq  \sp_{\rm a}(T),$
where $\partial \sp(T)$ denotes the topological boundary of
$\sp(T)$, and that $\sp_{\rm r}(T)\subseteq \sp_{\rm p}(T\s)$.
Also note that
$\sigma(T) = \sp_{\rm a}(T)\cup\sp_{\rm r}(T)$ and that the union is disjoint.

If $S$ and $T$ are bounded operators on $X$ satisfying 
$ST = TS$, then 
\begin{equation}\label{eq:cont-dep}
\delta(\sp(S),\sp(T))\le r(S-T),
\end{equation}
where $\delta(K,L)$ denotes the Hausdorff
distance between the compact sets $K$ and $L$ \cite[Theorem 3.4.1]{Au}.
We will apply this result in the following situation.
Let $(X_n)_{n\ge 0}$ be a sequence of nonzero complemented subspaces of $X$
such that $X_n\cap X_m = \{0\}$ for all $n,m\ge 0$ with $n\not=m$.
Let $(\pi_n)_{n\ge 0}$ be a corresponding sequence of projections. 
For each $n\ge 0$ let 
$P_n:=\bigoplus_{j=0}^n \pi_j$. 
Let $T$ be a bounded operator on $X$ which commutes
with each $P_n$. We define operators $T_n$ on $X$ and $S_n$ on $X_n$ by
$T_n:= T\circ P_n$ and $S_n:=T|_{X_n}.$  

\begin{proposition}\label{prop:direct_sum}
Under the above assumptions, if
\begin{equation}\label{eq:norm-conv_n}
\limn \n T - T_n \n = 0,
\end{equation}
then
\begin{equation}\label{eq:sp_union}
\sp (T)=\overline{ \bigcup_{n=0}^\infty \sigma (T_{n})}
=\overline{ \bigcup_{n=0}^\infty \sigma (S_{n})}.
\end{equation}
\end{proposition}
\begin{proof} 
Let $Y_n := \ker P_n  =  (I-P_n)X$. 
Then we have a direct sum decomposition
$X = X_0\oplus \dots\oplus X_n \oplus Y_{n}$, relative to which 
we have $T_n = S_0\oplus \dots\oplus S_n \oplus 0$. From this 
we easily infer that
$\sp (T_n)= \{ 0\}\cup \bigcup_{j=0}^n \sigma (S_j).$
Moreover, from $\lim_{n\to\infty}\n S_n\n =\lim_{n\to\infty}\n T_n - T_{n-1}\n 
= 0$ we obtain that 
$0\in\overline{\bigcup_{n\ge 0} \sp(S_n)}$.
The second identity in \eqref{eq:sp_union} immediately follows.
The inclusion `$\subseteq$' in the first identity follows from 
\eqref{eq:norm-conv_n} and \eqref{eq:cont-dep}, while the 
inclusion `$\supseteq$' follows from the obvious inclusions 
$\sp(T_n)\subseteq \sp(T)$.
\end{proof}

Let $X$ and $Y$ be Banach spaces and let
$X\widehat\otimes Y$ denote the completion of $X\otimes Y$ 
with respect to a uniform cross norm. If $S$ and $T$ are bounded
operators on $X$ and $Y$ respectively, then
the operator $S\otimes T$ defined on $X\otimes Y$
$$(S\otimes T)(x\otimes y):= (Sx\otimes Ty) $$
uniquely extends to a bounded operator $S\widehat\otimes T$
on $X\widehat\otimes Y$ of norm $\n S\widehat\otimes T\n \le \n S\n\,\n T\n$.
The spectrum of $S\widehat\otimes T$
is given by the following identity due to Schechter \cite{Sch}:
\begin{equation}\label{eq:Sch}
\displaystyle\sp(S\widehat\otimes T) = \{\eta\cdot\zeta: \
\eta\in\sp(S), \ \zeta\in\sp(T)\}.
\end{equation}
We will only need the Hilbert space version, which was obtained
earlier by Brown and Pearcy \cite{BP}.

\subsection{Reproducing kernel Hilbert spaces}
The material of this subsection is needed in Section \ref{sec5}.
Let $E$ be a {\em real} Banach space with dual $E\s$ and let
$Q\in{\mathscr L}(E\s,E)$
be a positive and symmetric linear operator,
i.e., we have $\lb Qx\s,x\s\rb\ge 0$ for all $x\s\in
E\s$ and $\lb Qx\s,y\s\rb = \lb Qy\s, x\s\rb$ for all $x\s,y\s\in E\s$.
On the range of $Q$, the bilinear mapping
$(Qx\s, Qy\s) \mapsto \lb Qx\s, y\s\rb$
defines an inner product. The completion of the range of $Q$
with respect to this inner product is a real Hilbert space $H_Q$, the
{\em reproducing kernel Hilbert space}
associated with $Q$.
The inclusion mapping from the range of $Q$
into $E$ extends to a continuous inclusion mapping $i_Q: H_Q\embed E$.
Upon identifying $H_Q$ and its dual in the canonical way
we have the operator identity
\begin{equation}\label{eq:fact} 
Q = i_Q \circ i_Q\s.
\end{equation}
If $Q_1, Q_2\in\calL(E\s,E)$ are positive and symmetric operators, then
we have $H_{Q_1}\subseteq H_{Q_2}$ as subsets of $E$ if and only if
there exists a constant $K\ge 0$ such that
\begin{equation}\label{eq:RKHS}
 \lb Q_1 x\s,x\s\rb \le K \lb Q_2 x\s,x\s\rb \qquad \hbox{for all}\  x\s\in E\s
\end{equation}
in which case the inclusion mapping $H_{Q_1}\embed H_{Q_2}$ is continuous.

If
$i:H\embed E$ is a continuous embedding of a real Hilbert space $H$
into $E$, then $Q:= i\circ i\s$ is positive and symmetric and
its reproducing kernel space $H_Q$ equals $H$. More precisely,
the mapping $i\s x\s \mapsto i_Q\s x\s$ defines an isometry from
$H$ onto $H_Q$ and from $i\circ i\s = Q =i_Q\circ i_Q\s$
we have $H=H_Q$ as subsets of $E$.

Examples of positive symmetric operators arise naturally in the theory of
Gaussian distributions. Recall that if
$\g$ is a centered Gaussian Radon measure on $E$, then
there exists a unique positive and symmetric operator $Q_\g\in \calL(E\s,E)$,
the {\em covariance operator} of $\g$, such that the Fourier transform of
$\g$ is given by
$$\widehat{\g}(x\s):= \int_E \exp(-i\lb x,x\s\rb)\,d\gamma(x)
= \exp(-\tfrac12\lb Q_\g x\s,x\s\rb) \qquad \hbox{for all}\ x\s\in E\s.$$
In this situation the reproducing kernel
Hilbert space $H_\g := H_{Q_\g}$ is separable,
the embedding $i_\g: H_\g\embed E$ is compact, and we have
$\g(\overline{H_\g}) = 1$, the closure being taken with respect
to the norm of $E$.

\section{The spectrum of second quantized operators}\label{sec3}

Let $H$ be a nonzero complex  Hilbert space. For $n\ge 0$ we let
$H^{\otimes n} = H\widehat\otimes\cdots \widehat\otimes H$
be the $n$-fold Hilbert tensor product
of $H$, with the understanding that
$H^{\otimes 0} =\C$. The Hilbert space direct sum
$$
\G(H):={\bigoplus}_{n\ge 0}  H^{\otimes n }
$$
is called the {\em Fock space} over $H$. 
The theory of Fock spaces is developed systematically in \cite{Pa}.

Given a bounded operator $T\in {\mathscr L}(H)$,
we use the notation
 $T^{\otimes n}=T\widehat\otimes\cdots\widehat\otimes T\in {\mathscr L}(H^{\otimes n})$,
with the understanding that
$T^{\otimes 0} = I$.
For later use we note that for all $S,T\in {\mathscr L}(H)$ and all $n\ge 1$
we have 
\begin{equation}\label{eq:Tn}
\n T^{\otimes n}\n_{\calL(H^{\otimes n})} = \n T\n^n
\end{equation}
and, by a simple telescoping argument,
\begin{equation}\label{eq:telesc}
 \n T^{\otimes n} - S^{\otimes n}\n_{\calL(H^{\otimes n})}
 \le \n T-S\n \cdot \sum_{j=0}^{n-1} \n S\n^j \n T\n^{n-1-j}.
\end{equation}
If $T$ is a contraction, the direct sum operator
$$
\Gamma( T):={\bigoplus}_{n\ge 0} T^{\otimes n}
$$
is well defined and defines a contraction on $\G(H)$.
This operator is called the {\em second quantization} of $T$.
We have the following algebraic relations:
\begin{equation}\label{eq:algebr}
\G(I)  = I, \qquad
\G(T_1 T_2) = \G(T_1)\G(T_2), \qquad
\G(T\s) =(\G(T))\s.
\end{equation}


\begin{proposition}\label{prop:sec_q} 
If $\n T\n < 1$, then 
$$
\sp (\Gamma(T))=\{1\}\cup\overline{\bigcup_{n\ge 1}\Bigl\{\prod_{j=1}^n   \l_j:  \
\l_j\in\sigma(T) ; \ j=1,\dots,n\Bigr\}}.
$$
\end{proposition}

\begin{proof}
Clearly, $\sigma(T^{\otimes 0}) = \{1\}$, whereas for $n\ge 1$ 
by repeated application of \eqref{eq:Sch} we have
\begin{equation}\label{eq:RS}
\sigma(T^{\otimes n})
=\Bigl\{\prod_{j=1}^n \l_j:  \ \l_j\in\sigma(T); \ j=1,\dots,n\Bigr \}.
\end{equation}
The result now follows from a straightforward application of
Proposition  \ref{prop:direct_sum}.
\end{proof}

For our applications in the next sections
we will be interested in the {\em symmetric Fock space} over $H$. 
This is the Hilbert space direct sum
$$ \Hs := {\bigoplus}_{n\ge 0} H^{\odot n},$$
where $H^{\odot n}$ denotes the closed subspace of $H^{\otimes n}$
spanned by all symmetric $n$-tensors, again with the understanding that
$H^{\odot 0} = \C$. If $T$ is a bounded operator on $H$, then $T^{\otimes n}$
maps $H^{\odot n}$ into itself. 
The restriction of $T^{\otimes n}$ to $H^{\odot n}$ will be denoted by 
$T^{\odot n}$.
If $T$ is a contraction, we define the {\em symmetric second quantization} of
$T$ by $$\Gs(T) := {\bigoplus}_{n\ge 0} T^{\odot n}.$$
Of course, $\Gs(T)$ is just the restriction of $\G(T)$ to $\Hs$.
The algebraic relations \eqref{eq:algebr} carry over in the obvious way.

Let $S_n$ denote the 
permutation group on $n$ elements. 
Given an element $h\in H$, the {\em creation operators}
$a_n^\dagger(h): H^{\odot n}\to H^{\odot(n+1)}$ are defined by
$$ 
\begin{aligned}
\ & a_n^\dagger(h)\sum_{\sigma\in S_n} g_{\sigma(1)}\otimes\dots
\otimes g_{\sigma(n)} 
\\ &  := \frac{1}{\sqrt{n+1}}
\sum_{\sigma\in S_n} \sum_{m=1}^{n+1} g_{\sigma(1)}
\otimes\dots\otimes g_{\sigma(m-1)}\otimes h\otimes g_{\sigma(m)}\otimes \dots \otimes g_{\sigma(n)},
\end{aligned}
$$
and the {\em annihilation operators}
$a_{n+1}(h): H^{\odot (n+1)}\to H^{\odot n}$ by
$$
\begin{aligned} 
\ & a_{n+1}(h)\sum_{\sigma\in S_{n+1}} g_{\sigma(1)}\otimes\dots
\otimes g_{\sigma(n+1)} 
\\ &  := \frac{1}{\sqrt{n+1}}
\sum_{\sigma\in S_{n+1}} \sum_{m=1}^{n+1} [g_{\sigma(m)},h]_H \ 
 g_{\sigma(1)}\otimes\dots\otimes g_{\sigma(m-1)}\otimes 
g_{\sigma(m+1)}\otimes\dots \otimes
g_{\sigma(n+1)}.
\end{aligned}
$$
These operators are well defined and bounded, and their operator norms
are bound\-ed by
\begin{equation}\label{eq:n-bound}
\n a_n^\dagger\n_{\calL(H^{\odot n}, H^{\odot(n+1)})} 
= \n a_{n+1}(h)\n_{\calL(H^{\odot (n+1)}, H^{\odot n})}  \le C_n \n h\n
\end{equation} 
with constants $C_n$ depending on $n$ only.
The first equality follows from the duality relations
\begin{equation}\label{eq:duality}
a_{n}^{\dagger *}(h) = a_{n+1}(h).\end{equation}  
Furthermore, we have  the commutation relations
\begin{equation}\label{eq:commut} 
 a_{n+2}(h) a_{n+1}^{\dagger}(h) -  a_{n}^{\dagger}(h)a_{n+1}(h)  =
\n h\n^2 I.
\end{equation}
For the proofs we refer to \cite{Pa}.
An obvious consequence of \eqref{eq:duality} and \eqref{eq:commut} is the lower bound
\begin{equation}\label{eq:lower-bound}
\n a_{n}^\dagger (h)g\n_{H^{\odot (n+1)}}^2
\ge \n g\n_{H^{\odot n}}^2 \n h\n^2.
\end{equation}

We will need the following result from the theory
of several complex variables, known as Hartog's theorem \cite[page 106]{Ha}, 
cf. also  \cite[Lemma 2]{RS}:
 
\begin{lemma}\label{lem:Hartog}
Let $K$ be a bounded set in $\C^n$, $n\ge 2$, and suppose that $f$
is analytic in a neighbourhood of $K$. If $f(p)=z$ for some point $p\in K$,
then there exists a point $p'\in\partial K$, the topological boundary of
$K$, such that $f(p')=z$.
\end{lemma}

We are now in a position  to prove the  following result.

\begin{theorem}\label{thm:equal-sp}
For all $n\ge 0$ we have
$$\sigma(T^{\odot n}) = \sigma(T^{\otimes n})=
\Bigl\{\prod_{j=1}^n \l_j:  \ \l_j\in\sigma(T); \ j=1,\dots,n\Bigr \}.
$$
\end{theorem}

In the proof below, and in the rest of the paper, we  
economize on brackets; for instance,
$T^{\odot n *}$ means $(T^{\odot n})\s$ and 
$T^{*\odot n}$ means $(T\s)^{\odot n}$.

\begin{proof}
The second equality has already been noted in \eqref{eq:RS}, 
so we concentrate on the proof that $\sigma(T^{\odot n}) = \sigma(T^{\otimes
n})$.
For $n=0$ this is trivial, so we fix $n\ge 1$.

In order to prove the inclusion 
 $\sigma(T^{\odot n})\subseteq
\sigma(T^{\otimes n})$ it suffices to check that  
$T^{\otimes n}$ maps $(H^{\odot n})^\perp$
into itself.  
For any elementary symmetric tensor $h\in H^{\odot n}$, say
$
h = \sum_{\sigma\in S_n} h_{\sigma(1)}\otimes\dots\otimes h_{\sigma(n)},
$
and any element $x\in H^{\otimes n}$ we have
\begin{equation}\label{eq:dual}
[h,T^{\otimes n}x]_{H^{\otimes n}} =  \Bigl[\sum_{\sigma\in S_n} T^*
h_{\sigma(1)}\otimes\dots\otimes T^*h_{\sigma(n)}, x\Bigr]_{H^{\otimes n}} 
 = [T^{*\otimes n}h,x]_{H^{\otimes n}}.
\end{equation}
Clearly, $H^{\odot n}$ is invariant under $T^{*\otimes n}$,
and therefore for $x\in (H^{\odot n})^\perp$ 
we obtain 
$[h,T^{\otimes n}x]_{H^{\otimes n}}= [T^{*\otimes n}h,x]_{H^{\otimes n}}=0.$
Thus, $T^{\otimes n}x\in (H^{\odot n})^\perp$. 
\medskip

Next we prove the inclusion
$\sigma(T^{\otimes n})\subseteq \sigma(T^{\odot n})$.
We proceed by induction on $n$, the case $n=1$ being trivial
since $T^{\otimes 1} = T^{\odot 1} = T$.
Assume that we already know that 
$\sigma(T^{\otimes n})\subseteq \sigma(T^{\odot n})$
and fix $\l\in \sigma(T^{\otimes (n+1)})$. 
We have to show that $\l\in  \sigma(T^{\odot (n+1)})$. 

Noting that $T^{\otimes (n+1)} = T^{\otimes n}\widehat\otimes T$,
by \eqref{eq:Sch} we have $\l=\zeta\cdot\eta$
with $\zeta\in \sigma(T^{\otimes n}) \subseteq \sigma(T^{\odot n})$
and $\eta\in \sigma(T)$.
By Lemma \ref{lem:Hartog} 
we may assume that $(\zeta,\eta) \in \partial(\sigma(T^{\odot
n})\times \sigma(T))$, 
so $\zeta\in\partial\sigma(T^{\odot n})$ or $\eta\in\partial\sigma(T)$.

{\em Case 1:} \ 
Assume that $\zeta\in\partial\sigma(T^{\odot n})$ and $\eta\in\sp_{{\rm a}}(T)$.
Since boundary spectrum belongs to the approximate point
spectrum we have  $\zeta\in\sp_{{\rm a}}(T^{\odot n})$. 
Let $(g_{k})_{k\ge 1} $ and 
$(h_{k})_{k\ge 1}$
be corresponding approximate eigenvectors
for $T^{\odot n}$ and $T$, respectively.
From $ T^{\odot (n+1)}a_{n}^\dagger(h_k)g_k = a_n^\dagger(Th_k)T^{\odot n}g_k$
we have 
$$
\begin{aligned}
\ & 
\n T^{\odot (n+1)}a_{n}^\dagger(h_k)g_k - \zeta\eta
\, a_{n}^\dagger(h_k)g_k\n_{H^{\odot(n+1)}}
\\ & \qquad \le 
 \n a_{n}^\dagger(Th_k)(T^{\odot n}g_k - \zeta g_k)\n_{H^{\odot(n+1)}}
+ |\zeta| \,\n a_{n}^\dagger(Th_k-\eta h_k)g_k \n_{H^{\odot(n+1)}}.
\end{aligned}
$$
Hence, by \eqref{eq:n-bound} and since by assumption we have 
$\n g_k\n_{H^{\odot n}}= \n h_k\n = 1$,
$$\lim_{k\to\infty} \n T^{\odot (n+1)} a_{n}^\dagger(h_k)g_k-
\zeta\eta \,a_{n}^\dagger(h_k)g_k\n_{H^{\odot(n+1)}} = 0.
$$
Moreover, by \eqref{eq:lower-bound},
$$\n a_{n}^\dagger(h_k)g_k\n_{H^{\odot(n+1)}} \ge \n g_k\n_{H^{\odot n}}\n h_k\n = 1.$$ 
Since by \eqref{eq:n-bound} the sequence $(a_{n}^\dagger(h_k)g_k)_{k\ge 1}$ 
is 
bounded, upon normalizing we obtain an approximate eigenvector
for $T^{\odot (n+1)}$
with approximate eigenvalue $\l= \zeta\cdot\eta$.

{\em Case 2:} \ Assume that $\zeta\in\partial\sigma(T^{\odot n})$ and 
$\eta\in\sp_{{\rm r}}(T)$.
Then also $\zeta\in\partial\sigma(T^{\odot n *})$ and hence
$\zeta\in\sp_{{\rm a}}(T^{\odot n *})=\sp_{{\rm a}}(T^{*\odot n})$. 
Also, $\eta\in\sp_{{\rm a}}(T^*)$,
and therefore $\l\in  \sp_{{\rm a}}(T^{*\odot (n+1)}) =\sp_{{\rm a}}(T^{\odot (n+1) *}) 
\subseteq \sigma(T^{\odot (n+1)})$ as in Case 1.
Here we used that for all $k\ge 1$ we have $T^{\odot k *} =  T^{* \odot k }$
by \eqref{eq:dual}.
 
{\em Case 3:} \ If  $\zeta\in\sp_{{\rm a}}(T^{\odot n})$ and
$\eta\in\partial\sigma(T)$ we proceed
as in Case 1.

{\em Case 4:} \ If $\zeta\in\sp_{{\rm r}}(T^{\odot n})$ and 
$\eta\in\partial\sigma(T)$ we proceed
as in Case 2.
\end{proof}

Arguing as in the proof of Proposition \ref{prop:sec_q} we now obtain:

\begin{theorem}\label{thm:sec-q}
If $\n T\n < 1$, then 
$$
\sp (\Gs(T))=\sp (\Gamma(T))
=\{1\} \cup \overline{\bigcup_{n\ge 1}\Bigl\{\prod_{j=1}^n  \l_j:  \ \l_j\in\sigma(T); \ j=1,\dots,n\Bigr\}}.
$$
\end{theorem}

If $T$ is  compact, this result simplifies as follows:

\begin{corollary}\label{cor:point-sp}
If $T$ is compact and satisfies $\n T\n<1$, then
$$
\sp (\Gs(T))=\sp (\Gamma(T))
=\{0\}\cup \{1\} \cup {\bigcup_{n\ge 1}\Bigl\{\prod_{j=1}^n  \l_j:  \ \l_j\in\sigma(T); \ j=1,\dots,n\Bigr\}}.
$$
\end{corollary}
\begin{proof}
We have already seen that $\sp (\Gs(T))=\sp (\Gamma(T))$;
for the second identity we have to show that 
$0\in \sp (\G(T))$ and to prove the inclusion 
`$\subseteq$'.

From $\n T\n<1$ and $\sp(T)\not=\emptyset$ 
we see that $0$ is in the closure of
$\bigcup_{n\ge 1}\bigl \{\prod_{j=1}^n   \l_j:  \
\l_j\in\sigma(T); \ j=1,\dots,n\bigr\}$. Hence, $0\in \sp (\G(T))$
by Proposition \ref{prop:sec_q}.

From \eqref{eq:Tn} and the compactness of $T^{\otimes n}$ we see that
$\G(T)$ is compact. Hence, $\sp(\G(T))\setminus\{0\} = \sp_{\rm p}(\G(T))\setminus\{0\}$.
Let $\l\in \sp(\G(T))\setminus\{0\}$ be arbitrary and choose an eigenvector $x$ for $\l$.
Then for all $n\ge 1$ we have 
$T^{\otimes n} P_n x = P_n \G(T)  x = \l \,P_n x,$
where $P_n$ denotes the orthogonal projection in $\G(H)$
onto $\bigoplus_{0\le j\le n} H^{\otimes j}$.
For sufficiently large $n$ we have $P_n x\not=0$, and for those $n$ we conclude
that $\l\in \sigma_{\rm p}(T^{\otimes n})$. In particular we see that
$\l = \prod_{j=1}^n  \l_j$ for certain $\l_j\in\sigma(T)$.
\end{proof}

\section{The $L^p$-spectrum of second quantized contraction semigroup 
generators}\label{sec4}

Throughout this section we fix an arbitrary real Banach space $E$
and a centered Gaussian Radon measure $\g$ on $E$.
Let $H_\g$ denote the reproducing kernel Hilbert space of $\g$ 
and let $i_\g: H_\gamma\embed E$ be the associated embedding.
Since $\g(\overline{H_\g})=1$, when considering the spaces
$L^p(E,\g)$ there will be no loss of generality
in assuming that $\g$ is {\em nondegenerate}, by which we mean that
$H_\g$ is dense in $E$.

Let $H:= H_{\g,\C}$ be the complexification of $H_\g$.
It is well known that the complex Hilbert space $L^2(E,\g)$ is canonically isometrically isomorphic
to the symmetric Fock space $\Gs(H)$.
We will describe this isometry briefly here;  for a more detailed discussion
we refer to  \cite{Ja}.
Each element $h\in H_\g$ of the form $h=i_\g\s x\s$ defines a real-valued
function $\phi_h\in L^2(E,\g)$ by
$ \phi_h(x) := \lb x,x\s\rb$
and we have 
$$ \n \phi_h\n_{L^2(E,\g)}^2 = \int_E \lb x,x\s\rb^2\,d\g(x) 
= \n i \s x\s\n_{H_\g}^2.
$$
Since $i_\g\s$ has dense range in $H_\g$, the mapping
$h\mapsto \phi_h$ uniquely extends to an isometry from $H_\g $ 
into the real part of $L^2(E,\g)$. By complexification we obtain an 
isometry $h\mapsto \phi_h$ from $H$
into $L^2(E,\g)$.
Using this isometry, for each $n\ge 1$ we define
 ${\H}_{\le n}$ as  the closed subspace of $L^2(E,\g)$ spanned by
 the constant one function ${\bf 1}$ and all 
products $\phi_{h_1}\cdot\hdots\cdot\phi_{h_m}$ of order $1\le m\le n$, where 
$h_1,\dots,h_n\in H$. 
We then let $\H_0 := \C{\bf 1}$ and define, for $n\ge 1$, the space
$\H_n$ as the orthogonal complement of 
$\H_{\le n-1}$ in $\H_{\le n}$. The complex form of the 
Wiener-It\^o decomposition theorem
asserts that we have an orthogonal direct sum decomposition
$$
L^2(E,\g)=\bigoplus_{n\ge 0}\H_n.
$$
The space $\H_n$ is usually referred to as the $n$-th {\em Wiener-It\^o chaos}.
Denoting by 
$I_n$ the orthogonal projection in $L^2(E,\g)$ onto $\H_n$,
it is not difficult to show that
$$
\langle I_n(\phi_{h_1}\dots\phi_{h_n}),I_n(\phi_{k_1}\dots\phi_{
k_n})\rangle =\sum_{\sigma\in S_n}[h_1,k_{\sigma (1)}]_H\dots
[h_n,k_{\sigma (n)}]_H.
$$
This shows that $\H_n$ is canonically isometric to $H^{\odot n}$
as a Hilbert space,
the isometry being given explicitly by
$$ I_n(\phi_{h_1}\cdot\dots\cdot\phi_{h_n}) \mapsto \frac1{\sqrt{n!}} \sum_{\sigma\in S_n}
h_{\sigma(1)}\otimes\dots\otimes h_{\sigma(n)}.
$$
Thus the Wiener-It\^o decomposition induces
a canonical isometry of $L^2(E,\g)$ and the symmetric Fock space
$\G^{\odot}(H)$.
 
Let us now assume that $A$ is the infinitesimal generator 
of a strongly continuous semigroup of contractions $S=\{S(t)\}_{t\ge 0}$
on $H$. 
We denote by $P_2=\{P_2(t)\}_{t\ge 0}$ its symmetric second quantization:
$$ P_2(t) = \Gs(S(t)).$$
By the isometry just described, $P_2$
induces a semigroup of contractions, also denoted by $P_2$, on
$L^2(E,\g)$.
Since $P_2$ is strongly continuous on each $\H_n$, it follows that
$P_2$ is strongly continuous on $L^2(E,\g)$. In fact, $P_2$ is doubly Markovian
and therefore $P_2$ extends uniquely to a strongly continuous
semigroup of contractions on $L^p(E,\g)$ for every $p\in [1,\infty)$; cf.
\cite{Si}. 

\begin{lemma}\label{lem:norm-cont} 
If $t\mapsto S (t)$ is norm continuous for $t>t_0$
and $S (t)$ is a strict contraction for $t>s_0$, then
for every $p\in (1,\infty)$, 
$t\mapsto P_p(t)$ is norm continuous for 
$t>\max\{t_0,s_0\}$.
\end{lemma}
\begin{proof}
First we consider the case $p=2$.
By  \eqref{eq:telesc}, for each $n\ge 1$ the restriction 
$P_{2,n}$ of $P_2$ to $\H_n$
is contractive and norm continuous for $t>t_0$.
For $t >s_0$, by \eqref{eq:Tn}  
we have 
$ P_2(t) = \sum_{n\ge 0} P_{2,n}(t)$
with convergence in the operator norm, uniformly on $[s,\infty)$ for every
$s>s_0$. Hence, for $t>\max\{t_0,s_0\}$ the
function $t\mapsto P_2(t)$ is  norm continuous, since on this interval
it is the locally uniform limit
of a sequence of norm continuous functions.

Next we take $p\in (1,2)$ and use the fact that 
$\n P_1 (t)-P_1 (s)\n_{\calL(L^1 (E,\g))}\le 2$
and the Riesz-Thorin interpolation theorem to find that
$$ \n P_p (t)-P_p (s)\n_{\calL(L_p (E,\g))}\le 2^{1-\theta_p}
\n P_2(t)-P_2(s)\n_{\calL(L^2(E,\g))}^{\theta_p}$$
where $(1-\theta_p)+\tfrac12\theta_p = \tfrac1p$.
For $p\in (2,\infty)$ we proceed similarly, 
this time interpolating between $L^2(E,\g)$ and $L^{p'}(E,\g)$ 
with $p'\in (p,\infty)$.
\end{proof}

In the next lemma we need some further 
results about the spaces $L^p(E,\g)$. We refer to \cite{Ja} for 
the proofs, which are based on standard hypercontractivity arguments.
For all $p\in (1,\infty)$ and $n\ge 0$ we have
$\H_n\subseteq L^p(E,\g)$ and the restrictions of the $L^2(E,\g)$-norm and the
$L^p(E,\g)$-norm are equivalent on $\H_n$.
Furthermore, the orthogonal projections $I_n$ in $L^2(E,\g)$
onto $\H_n$  extend uniquely  to projections $I_{p,n}$
in $L^p(E,\g)$ onto $\H_{n}$.
As a subspace of $L^p(E,\g)$, $\H_n$ will be denoted by $\H_{p,n}$.
By the observations just made, each $\H_{p,n}$ is complemented in $L^p(E,\g)$.

\begin{lemma}\label{lem:restr-Lp}
For all $p\in (1,\infty)$ there is a constant $\theta_p\in (0,1]$
such that for all $n\ge 1$ and $t\ge 0$ we have
$$\n P_{p,n}(t)\n_{\calL(\H_{p,n})}\le \n
S(t)\n^{n\theta_p}.$$
As a consequence we have
$P_p(t) = \sum_{n\ge 0} P_{p,n}(t),$
the convergence being in the operator norm, uniformly on $[t_0,\infty)$ for all $t_0>0$.
\end{lemma}
\begin{proof}
Fix $t\ge 0$. For $p=2$ we may take $\theta_p =1$.
For $p\in (1,2)$, choose $p'\in (1,p)$ and recall that $P_{p'}$ is
a contraction semigroup on $L^{p'}(E,\g)$.
In particular, by taking restrictions, we see that
$ \n P_{p',n}(t)\n_{\calL(\H_{p',n})}\le 1$ for all $n\ge 1$.
Since we also have
$\n P_{2,n}(t)\n_{\calL(\H_n)}= \n S (t)\n^n,$
for $p\in (1,2)$ the result now follows by
interpolation; notice that
$(\H_{n},\H_{p',n})_{\theta_p} = \H_{p,n}$ with
$\tfrac1{p'}(1-\theta_p)+\tfrac12 \theta_p = \tfrac1p$.
For $p\in (2,\infty)$ we proceed similarly, this time
interpolation between $\H_{n}$ and $\H_{p',n}$ with $p'\in (p,\infty)$.

By the estimate just proved, the series 
$\sum_{n\ge 0} P_{p,n}(t)$
converges in the operator norm, uniformly on $[t_0,\infty)$ for every $t_0>0$.
Moreover, on the dense subspace spanned by the spaces $\H_{p,n}$
the sum equals $P_p(t)$. This completes the proof.
\end{proof}

The infinitesimal generators of the semigroups $P_p$ 
and $P_{p,n}$ will be denoted by $L_p$ and $L_{p,n}$, respectively.

In the proof of the main result of this section, Theorem \ref{thm:sp-L_p} below,
we shall use the well-known fact \cite[Theorem II.4.18]{EN} that 
if $A$ is the infinitesimal generator of a strongly continuous and 
eventually norm continuous semigroup on $X$,
then 
\begin{equation}\label{eq:sp-norm-cont} 
\hbox{for every $a\in\R$ the set
$\{\l\in\C: \ \l\in\sigma(A), \ \Re\l\ge a\}$ is bounded.}
\end{equation}

Let us call a semigroup $T$ 
{\em strictly contractive} if $\n T(t)\n < 1$ for all $t>0$.

\begin{theorem}\label{thm:sp-L_p}  
Let $p\in (1,\infty)$.
If $S$ is strictly contractive and eventually norm continuous, then 

\begin{equation}\label{eq:sp_L2-n}
\begin{aligned}
\  \sp (L_{p,0})  = \{0\},\quad
   \sp (L_{p,n})  = {\Bigl\{\sum_{j=1}^n  \zeta_j:  
\ \zeta_j\in\sigma(A); \ j=1,\dots,n\Bigr\}},
\end{aligned}
\end{equation}
and
\begin{equation}\label{eq:sp_L2}
\begin{aligned}
 \sp (L_p) & = \{0\} \cup \overline{\bigcup_{n\ge 1}\Bigl\{\sum_{j=1}^n  \zeta_j:  
\  \zeta_j\in\sigma(A); \ j=1,\dots,n\Bigr\}}
\\ & =
\overline{\Bigl\{ \sum_{j=1}^n k_j z_j: \ \ k_j\in\N, \ z_j
\in\sp(A);\ j=1,\dots,n; \ n\ge 1\Bigr\}}.
\end{aligned} 
\end{equation}
\end{theorem}

\begin{proof} Let $p\in (1,\infty)$ be fixed.
It is clear that $\sp (L_{p,0})  = \{0\}$, so let us
fix $n\ge 1$. 

From Theorem \ref{thm:equal-sp}, Lemma \ref{lem:norm-cont}, and the spectral mapping
theorem for eventually norm continuous semigroups, first applied to
$P_{p,n}$ and then to $S$, for $t\ge 0$ we obtain  
\begin{equation}\label{eq:Sp-Lpn}
\begin{aligned}
\exp\bigl(t\,\sp(L_{p,n})\bigr)
 & = \sigma(P_{p,n}(t))\setminus\{0\}
\\ & =  {\Bigl\{\prod_{j=1}^n  \l_j:  
\ \l_j\in\sigma(S(t))\setminus\{0\} ; \ j=1,\dots,n\Bigr\}}
\\ & =  {\Bigl\{\prod_{j=1}^n \exp({t\zeta_j}):  
\ \zeta_j\in\sigma(A) ; \ j=1,\dots,n\Bigr\}}
\\ & = {\Bigl\{\exp\Bigl({\,t\sum_{j=1}^n\zeta_j}\Bigr) :  
\ \zeta_j\in\sigma(A) ; \ j=1,\dots,n\Bigr\}}.
\end{aligned}
\end{equation}
In order to obtain the corresponding equality for $\sp(L_p)$ we 
first check that for all $t\ge 0$,
$$ 
\sp (P_p(t)) 
=\{1\} \cup \overline{\bigcup_{n\ge 1}\Bigl\{\prod_{j=1}^n  \l_j: 
 \ \l_j\in\sigma(S(t)); \ j=1,\dots,n\Bigr\}}.
$$
Clearly this holds for $t=0$, and for fixed $t>0$ this follows from Proposition 
\ref{prop:direct_sum}, Theorem \ref{thm:equal-sp}, the expansion
in Lemma \ref{lem:restr-Lp}, 
and the fact that
$\H_n = \H_{p,n}$ with equivalent norms.
Repeating the argument of \eqref{eq:Sp-Lpn} we obtain
\begin{equation}\label{exp}
\exp\bigl(t\,\sp(L_{p})\bigr)
 =\{1\}\cup \overline{\bigcup_{n\ge
 1}\Bigl\{\exp\Bigl({\,t\sum_{j=1}^n\zeta_j}\Bigr) :  
\ \zeta_j\in\sigma(A); \ j=1,\dots,n\Bigr\}}\setminus\{0\}.
\end{equation}
Let 
$$B_n =  \Bigl\{ {\sum_{j=1}^n\zeta_j} :  
\ \zeta_j\in\sigma(A); \ j=1,\dots,n\Bigr\}\qquad (n\ge 1)
$$
and $$B :=  \bigcup_{n\ge 1}\Bigl\{{\sum_{j=1}^n\zeta_j} :  
\ \zeta_j\in\sigma(A); \ j=1,\dots,n\Bigr\}.
$$
For \eqref{eq:sp_L2-n} we have to 
  prove the  inclusions ${B_n}\subseteq \sp(L_{p,n})$
and $\sp(L_{p,n})\subseteq {B_n}$;
 for \eqref{eq:sp_L2} we have to 
  prove the  inclusions $\overline{B}\subseteq \sp(L_{p})$
  (the inclusion $\{0\}\subseteq \sp(L)$ being trivial)
and $\sp(L_{p})\setminus\{0\}\subseteq \overline{B}$.
We shall prove the latter two; the former two are proved in the same way.
We adapt an argument from \cite{He}.

\smallskip
$\bullet$ \ $ \overline{B}\subseteq \sp(L_p)$: \ \
\smallskip
 
Since $\sp(L_p)$ is closed, it suffices to prove that 
${B}\subseteq \sp(L_p)$. Fix an arbitrary $\zeta\in{B}$,
say
$\zeta= {\sum_{j=1}^n\zeta_j}$ with $n\ge 1$ and 
$\zeta_j\in\sigma(A)$ $(j=1,\dots,n)$.
By \eqref{exp}, for every $t>0$ we find an element $\zeta(t)\in\sp(L_p)$ and 
integer $N(t)\in\Z$
such that 
\begin{equation}\label{eq:mu}
\zeta= 2\pi i t^{-1} N(t)+\zeta(t).
\end{equation}
From $\Re\zeta(t)=\Re\zeta $ and  \eqref{eq:sp-norm-cont} we see
that there is a constant $C\ge 0$
such that $|\Im\zeta(t)|\le C$ for all $t>0$.
Comparing imaginary parts in \eqref{eq:mu} and letting $t\downarrow 0$
we see that $N(t)=0$ for small enough 
$t$. For those $t$ we then have $\zeta = \zeta(t) \in\sp(L_p)$.

\smallskip
$\bullet$ \ $\sp(L_p)\setminus\{0\}\subseteq \overline{B}$: \ \
\smallskip
 
The proof proceeds along the same lines, but extra care is needed
to control the number of terms occurring in the sums defining the elements
of $B$.
Fix $ z\in\sigma(L_p)\setminus\{0\}$.
Since $\n S(t)\n<1$ for all $t>0$, standard arguments from semigroup theory 
imply that $S$ is uniformly exponentially stable. 
Choose $\omega>0$ and $M\ge 1$ such that $\n S(t)\n \le Me^{-\omega t}$
for all $t\ge 0$. Then, $\sigma(A)\subseteq\{\lambda\in \C: \ \Re\lambda\le -\omega\}$.
Together with \eqref{exp} this implies that $\Re  z\le -\omega$.
By \eqref{eq:sp-norm-cont}, there exists a constant $C\ge 1$
such that for every $w\in \sigma(A)$ satisfying $\Re w\ge 2\Re z$
we have $|\Im w|\le C$. In particular,
\begin{equation}\label{eq:Imz}
 |\Im z| \le C.
\end{equation}
For reasons that will become clear soon we fix $t_0>0$
subject to the condition that
$$
2\pi t_0^{-1}  > C\bigl(1 + \tfrac{1}{\omega}|\Re z|\bigr).
$$
Choose $0<\e<1$ so small that
\begin{equation}\label{eq:eps}
2\pi t_0^{-1}  \ge C\bigl(1 +  (\tfrac{1+\e}{\omega}+\e)|\Re z|\bigr).
\end{equation}
Fix $0<t<t_0$ arbitrary. By \eqref{exp}
there exists a sequence of complex numbers $( z_k(t))_{k\ge 1}$ such that 
\begin{equation}\label{eq:convz}
\lim_{k\to\infty} z_k(t) =  z
\end{equation} 
with $ \exp\bigl(t z_k(t)) \in \exp(tB) $ for all $k$. Most of the remaining
argument is devoted to proving that
$ z_k(t)\in B$ for all sufficiently large $k$.

Choose $k_0(t)$ so large that 
\begin{equation}\label{eq:k_0}
| z_k(t)- z|\le \e|\Re  z|
\end{equation}
for all $k\ge k_0(t)$.
Fix $k\ge k_0(t)$ and notice that, by \eqref{eq:k_0} and the fact that
$0<\e<1$,
\begin{equation}\label{eq:k_0t}
\Re  z_k(t)\ge 2\Re z.
\end{equation}
Choose   integers
$n_k(t)\ge 1$ and $N_k(t)\in \Z$ such that
\begin{equation}\label{eq:repr-mu} 
 z_k(t) = 2\pi i t^{-1}\, N_k(t)+ \sum_{j=1}^{n_k(t)}\zeta_{j,k}(t)
\end{equation}
with all $\zeta_{j,k}(t)$ in $\sigma(A)$. Note that for all $j$,
\begin{equation}\label{eq:repr-mu2} 
\Re\zeta_{j,k}(t) \le -\omega
\end{equation} and hence, by \eqref{eq:repr-mu},
\begin{equation}\label{eq:Re1}
\Re  z_k(t) \le -n_k(t)\omega.
\end{equation}
Also notice that from \eqref{eq:k_0t}, \eqref{eq:repr-mu}, and
 \eqref{eq:repr-mu2},
\begin{equation}\label{eq:repr-mu3} 
\Re\zeta_{j,k}(t) \ge 2\Re z.
\end{equation}
From \eqref{eq:k_0} and \eqref{eq:Re1} we deduce that
$$
n_k(t) \omega \le |\Re  z_k(t)|  \le (1+\e)|\Re z| .$$ 
By \eqref{eq:repr-mu3} and the choice of $C$ we have  
$|\Im\zeta_{j,k}(t)|\le C$ and therefore, by 
\eqref{eq:Imz}, \eqref{eq:k_0}, \eqref{eq:repr-mu}, \eqref{eq:repr-mu3}, 
and the fact that $C\ge 1$,
\begin{equation}\label{eq:contrad1}
\begin{aligned}
C \ge |\Im z|
& \ge |\Im z_k|-\e C|\Re z|
\\ & \ge 2\pi t^{-1}\,N_k(t)   - Cn_k(t) -\e C|\Re z| 
\\ & \ge 2\pi t^{-1}\,N_k(t)   - (1+\e)C|\Re z|/\omega
-\e C|\Re z|.
\end{aligned}
\end{equation}
If $N_k(t)$ were nonzero, then $N_k(t)\ge 1$
and in view of $0<t<t_0$ 
the right hand side would be strictly greater than 
\begin{equation}\label{eq:contrad2} 
2\pi t_0^{-1}  - (1+\e)C|\Re z|/\omega
-\e C|\Re z| \ge  C,
\end{equation}
where the inequality follows from the choice of $\e$ in \eqref{eq:eps}.
By comparing  \eqref{eq:contrad1} and \eqref{eq:contrad2}  
we see that we have arrived at a 
contradiction. Thus, $N_k(t)=0$ and therefore,
$ z_{k}(t) = \sum_{j=1}^{n_k(t)}\zeta_{j,k}(t).$
It follows that $ z_{k}(t)\in B.$
 
So far, $k\ge k_0(t)$ was fixed. By letting $k\to\infty$ and recalling
\eqref{eq:convz} we obtain  $ z\in \overline{B}$.
\end{proof}

If the eventual norm continuity assumption is strengthened to
eventual compactness, there is no need to
take the closure in \eqref{eq:sp_L2}. This is the content
of the following semigroup analogue of Corollary \ref{cor:point-sp}:

\begin{corollary}\label{cor:Lp-point-sp}
Let $p\in (1,\infty)$. If $S$ is strictly contractive and 
eventually compact, then $P_p$ is eventually compact and 
$$\sigma(L_p) =
{\Bigl\{ \sum_{j=1}^n k_j z_j: \ \ k_j\in\N, \ z_j
\in\sp(A);\ j=1,\dots,n; \ n\ge 1\Bigr\}}.
$$
\end{corollary}
\begin{proof}
Let $S(t)$ be compact for $t>t_0$.
Since $\H_n = \H_{p,n}$ with equivalent norms and since $T^{\odot n}$ is compact
whenever $T$ is, the operators $P_{p,n}(t)$ are compact for $t>t_0$.
The expansion in Lemma \ref{lem:restr-Lp} then shows that $P_p(t)$ is
compact for $t>t_0$. Hence by the spectral mapping theorem for the point spectrum, 
$\sigma(L_p)$ consists of isolated eigenvalues and the result follows from
 Theorem  \ref{thm:sp-L_p}.
\end{proof}

\section{The $L^p$-spectrum of Ornstein-Uhlenbeck operators}
\label{sec5}

Let $E$ be a real Banach space, let $A$ be infinitesimal generator
of a strongly continuous semigroup 
$S=\{S(t)\}_{t\ge 0}$   on $E$, and let
$Q\in\calL(E\s,E)$ be a positive and symmetric operator.
In this section we shall apply our abstract results to the 
Ornstein-Uhlenbeck operator $L$, given on a suitable
core of cylindrical functions by
\begin{equation}\label{eq:defL}
 Lf(x) := \tfrac12 \hbox{Tr}\, (QD^2 f(x)) + \lb Ax, D f\rb\qquad(x\in E)
\end{equation}
where $D$  denotes the Fr\'echet derivative. The operator $L$ arises as
the infinitesimal generator of the transition semigroup of the 
Markov process $\{U_x(t)\}_{t\ge 0}$ that solves the stochastic linear
evolution equation
\begin{equation}\label{sACP}
\begin{aligned}
dU(t) & = AU(t)\,dt + dW_Q(t) \qquad (t\ge 0) \\
 U(0) & = x 
\end{aligned}
\end{equation}
where $W_Q$ is a cylindrical $Q$-Wiener process in $E$; cf. \cite{BN, DZ}.

For $t>0$ we define the positive symmetric operators
$Q_t\in\calL(E\s,E)$ by
$$ Q_t x\s := \int_0^t S(s)QS\s(s)x\s\,ds\qquad (x\s\in E\s). $$
The right hand side integral is
easily shown to exist as a Bochner integral in $E$.
In order to give a rigorous description of the operator $L$,
unless otherwise stated
we shall assume in the remainder of
this section that the following two hypotheses hold:

\begin{quote}
\begin{enumerate}
\item[(H$Q_\infty$)]
The weak operator limit $Q_\infty := \lim_{t\to\infty}Q_t$
exists in $\calL(E\s,E)$;
\item[(H$\mu$)\ \ ]
The operator $Q_\infty$ is the covariance of a centered Gaussian Radon measure
$\mu$ on $E$.
\end{enumerate}
\end{quote}
It will follow from Lemma \ref{lem:Hinfty} below that there is no loss of generality in assuming that
$\mu$ is nondegenerate.

Some comments are in order.
\begin{enumerate}
\item By
Hypothesis (H$Q_\infty$), for all $x\s,y\s\in E\s$ we have
$$ \lb Q_\infty x\s ,y\s\rb = \int_0^\infty \lb S(s)QS\s(s)x\s, y\s\rb\,ds,$$
the scalar integrals being defined in the improper sense. The positivity of
$Q$ together with a polarization argument imply that these integrals are
actually absolutely convergent.
\item
Hypothesis (H$\mu$) and a
standard tightness argument imply that
each $Q_t$ is the covariance operator of a
centered Gaussian Radon measure $\mu_t$, and we have $\lim_{t\to\infty}\mu_t =
\mu$ weakly.
\item If $E$ is separable, the word `Radon' may be replaced by `Borel'.
\item
If $E$ is a Hilbert space and we identify $E$ and $E\s$ in the usual way,
then Hypotheses (H$Q_\infty$) and (H$\mu$) hold if and only if each $Q_t$ is of trace class and  
$\sup_{t>0} \, \hbox{Tr}\,Q_t < \infty.$
Furthermore if Hypothesis (H$Q_\infty$) holds, then Hypothesis (H$\mu$) holds if and only if $Q_\infty$ is 
of trace class. In particular if $\dim E<\infty$, then Hypothesis (H$Q_\infty$) implies (H$\mu$).
\end{enumerate}
As is shown in \cite{BN, DZ},
the existence of the measures $\mu_t$ is equivalent to the existence of
a (necessarily unique) weak solution $\{U(t,x)\}_{t\ge 0}$ of \eqref{sACP}.
This solution is Gaussian; the random variable $U(t,x)$ has mean $S(t)x$ and
covariance operator $Q_t$. The solution is also Markovian
and its transition semigroup $P=\{P(t)\}_{t\ge 0}$ on $B_b(E)$
is given by
$$P(t)f(x) = {\mathbb E}\, f(U(t,x)) = \int_E f(S(t)x+y)\,d\mu_t(y) \qquad
(x\in E, \ f\in B_b(E)).$$
Here $B_b(E)$  denotes the space of
bounded complex-valued Borel measurable functions on $E$.
This semigroup leaves $C_b(E)$, the subspace of all continuous functions in
$B_b(E)$, invariant. Although the restricted semigroup generally fails to be
strongly continuous in the norm topology of $C_b(E)$, it
is strongly continuous in the strict topology of $C_b(E)$. This is, by
definition, the finest locally convex topology $\tau$
on $C_b(E)$ that agrees with the
compact-open topology on bounded sets. As a result,
the infinitesimal generator $L$ of $P$ is well defined as a linear operator
on the domain
$${\mathscr D}(L) = \Bigl\{f\in C_b(E): \ \hbox{$\tau$-}
\lim_{t\downarrow 0} \frac1t (P(t)f-f)\
\hbox{exists in $C_b(E)$}\Bigr\}.
$$
On a suitable core of cylindrical functions,  $L$ is given by \eqref{eq:defL}.
The measure $\mu$ is {\em invariant} for $L$
in the sense that for all $t\ge 0$ and $f\in B_b(E)$ we have
$$\int_E P(t)f(x)\,d\mu(x) = \int_E f(x)\,d\mu(x). $$
For more details we refer to the survey paper \cite{GN}.

By standard arguments, the invariance of $\mu$ implies that $P$
extends to a strongly continuous
contraction semigroup $P_p$ on $L^p(E,\mu)$ for all $p\in [1,\infty)$.
The infinitesimal generator of $P_p$ will be denoted by $L_p$. In order to establish the
relationship between these semigroups and the ones studied in the previous
section, we describe next how $P_2$ arises as a second quantized semigroup.

Let us denote the reproducing kernel Hilbert space associated with $Q_\infty$
by $H_\mu$ and the
embedding $H_{\mu}\embed E$  by $i_\mu$. By \eqref{eq:fact} we have
$Q_\infty = i_\mu\circ i_\mu\s$.
The key fact is the following result, due to
Chojnowska-Michalik and Goldys \cite {CG}
under some additional assumptions; the present formulation was given in
 \cite[Theorem 6.2]{Ne}.

\begin{lemma}\label{lem:Hinfty} Assume Hypothesis {\rm (H$Q_\infty$)}.
The space $H_\mu$
is invariant under the action of $S$,
and the restriction of $S$ to $H_\mu$, denoted by $S_\mu$,
is a strongly continuous semigroup of contractions
on $H_\mu$.
\end{lemma}
By complexification, $S_{\mu,\C}$ is a
strongly continuous semigroup of contractions on
$H_{\mu,\C}$, and upon
identifying $L^2(E,\mu)$ with $\Gs(H_{\mu,\C})$
as explained
in the previous section we have the following representation of $P_2$, again
due to Chojnowska-Michalik and Goldys \cite {CG}; see
also \cite[Theorem 6.12]{Ne}:

\begin{proposition}\label{prop:sec-q}
For all $t\ge 0$ we have
$P_2(t) = \Gs(S_{\mu,\C}\s(t)).$
\end{proposition}

Under this identification, the semigroup $P_p$ agrees with the one introduced in
the previous section.
As an immediate consequence of  Theorem \ref{thm:sp-L_p},
Corollary \ref{cor:Lp-point-sp}, and Proposition
\ref{prop:sec-q}
we obtain:

\begin{theorem}\label{thm:sp-OU}
If $S_\mu$ is strictly contractive and eventually norm continuous, then
$$
\begin{aligned}
 \sp (L_p) =
\overline{\Bigl\{ \sum_{j=1}^n k_j z_j: \ \ k_j\in\N, \ z_j
\in\sp(A_\mu);\ j=1,\dots,n; \ n\ge 1\Bigr\}}.
\end{aligned}
$$
If in addition $S_\mu$ is compact, then
$$\sigma(L_p) =
{\Bigl\{ \sum_{j=1}^n k_j z_j: \ \ k_j\in\N, \ z_j
\in\sp(A_\mu);\ j=1,\dots,n; \ n\ge 1\Bigr\}}.
$$
\end{theorem}

Notice that $\sp(L_p)$ depends on $\sp(A_\mu)$ rather than on $\sp(A)$.
This can be understood by observing that the definition of $L_p$ depends not
only on $A$, but also on $Q$.
On a deeper level, the abstract results of the previous section show that Theorem \ref{thm:sp-OU}
is in fact completely natural: Theorem \ref{thm:sp-L_p} shows that it
can be interpreted as saying that $\sp(L_p)$ can be
computed from the part of $L_p$ in the first Wiener chaos.
Of course, this limits that practical use of Theorem \ref{thm:sp-OU} to some
extent, as in general it may be difficult to compute $\sp(A_\mu)$
from $A$ and $Q$.

In the next two subsections we prove that the assumptions on $S_\mu$
are automatically satisfied
in two important cases: the strong Feller case and the finite-dimensional
case. In both cases we check the strict contractivity assumption
by an appeal to the following result, due to
Chojnowska-Michalik and Goldys \cite {CG}; cf. also \cite[Theorem 6.3]{Ne}.

\begin{lemma}\label{lem:str-contr}
Assume Hypothesis {\rm (H$Q_\infty$)} and fix $t>0$. Then $\n
S_\mu(t)\n_{\calL(H_\mu)}<1$ if and only if
$H_t = H_\mu$ with equivalent norms.
\end{lemma}
Here $H_t$ is the reproducing kernel Hilbert space associated with
the positive symmetric operator $Q_t$ introduced in the previous section.

The following example shows that a uniformly exponentially
stable semigroup may fail to be strictly contractive even if $\dim E<\infty$,
and that even if $S_\mu$
is strictly contractive it may happen that there exists no constant $a>0$
such that $\n S_\mu(t)\n_{\calL(H_\mu)} \le e^{-a t}$
for all $t\ge 0$.

\begin{example} For $\omega>0$ we consider the semigroup
$S^{(\omega)}$ on $E=\R^2$
defined by $S^{(\omega)}(t) = \displaystyle  e^{-\omega t}\Bigl(
\begin{array}{cc}
1 & t \\ 0 & 1
\end{array}\Bigr).
$
This semigroup is uniformly exponentially
stable, but for each $0<\omega<1$ we have
$\n  S^{(\omega)}(t)\n > 1$ for $t>0$ small enough.

Let us now take
$\displaystyle Q=\Bigl(
\begin{array}{cc}
0 & 0 \\ 0 & 1
\end{array}\Bigr).
$ With this choice, the limit $Q_\infty^{(\omega)}=\lim_{t\to\infty}
Q_t^{(\omega)}$
exists for every $\omega>0$. Let $\mu^{(\omega)}$ be the corresponding Gaussian
measures. Taking $\omega=1$ and $\mu:=\mu^{(1)}$, by the computations in 
\cite[Example 5.5]{GN} we have
$$\n S_\mu(t)\n_{\calL(H_\mu)} = e^{-t}\bigl(t+\sqrt{t^2+1}\bigr).$$
Thus, $ S_\mu(t)$ is strictly contractive on $H_\mu$ but there
exists no $a>0$ such that $\n S_\mu(t)\n_{\calL(H_\mu)} \le e^{-a t}$
for all $t\ge 0$.
\end{example}

\subsection{The strong Feller case}
 \label{sec6}
Throughout this subsection we assume Hypotheses (H$Q_\infty$) and (H$\mu$).
The transition semigroup $P$ is called {\em strongly Feller}
if
$P(t)f\in C_b(E)$ for all $f\in B_b(E)$ and $t>0$.
We have the following
reproducing kernel Hilbert space characterization of this property; see
\cite{DZ} and \cite[Corollary 2.3]{Ne}.

\begin{lemma}\label{lem:Feller}
The transition semigroup $P$ is strongly Feller if and only if
$S(t)E\subseteq H_t$ for all $t>0$.
\end{lemma}

The assumptions on $S_\mu$ in Theorem \ref{thm:sp-OU} are satisfied in the strong Feller case:

\begin{proposition}
If the transition semigroup $P$
is strongly Feller, then $S_\mu$ is compact and strictly contractive.
\end{proposition}

\begin{proof}
By \eqref{eq:RKHS} we have a continuous inclusion $H_t\embed H_\mu$; the
inclusion mapping will be denoted by $i_{t,\mu}$.
Denoting the inclusion mapping $H_t\embed E$ by $i_{t}$,
we have
$i_t = i_\mu\circ i_{t,\mu}.$
By Lemma \ref{lem:Feller} we have a factorization
\begin{equation}\label{eq:S_infty}
S_\mu(t) =  i_{t,\mu}\circ \Sigma(t)\circ i_\mu,
\end{equation}
where
$\Sigma(t)$ is the operator $S(t)$, viewed as an operator
from $E$ into $H_t$.
Recalling from Section \ref{sec2} that
$i_\mu$ is compact, it
follows that  $S_\mu(t)$ is compact for every $t>0$.

By Lemma \ref{lem:str-contr}  it remains to prove
 that for all $t>0$ we have $H_t = H_\mu$ with equivalent
norms. We have already seen that $H_t \embed H_\mu$ with
continuous inclusion.
To obtain the reverse inclusion we apply $i_\mu$ on both sides of 
\eqref{eq:S_infty} to obtain the identity
$$
S(t)\circ i_\mu = i_\mu\circ S_\mu(t) =
 i_{t}\circ \Sigma(t) \circ i_\mu .
 $$
Together with the  identity $Q_\infty = Q_t + S(t)Q_\infty S\s(t)$, which
follows directly from the definitions of $Q_t$ and $Q_\infty$,
for $x\s\in E\s$ we obtain
$$
\begin{aligned}
 \lb Q_\infty x\s,x\s\rb
  &  = \lb Q_t x\s, x\s\rb + \lb Q_\infty S\s(t) x\s, S\s(t) x\s\rb
\\  &  = \lb Q_t x\s, x\s\rb + [i_\mu\s S\s(t) x\s, i_\mu\s S\s(t) x\s]_{H_\mu}
\\ &  = \lb Q_t x\s, x\s\rb
 + [i_\mu\s  \Sigma\s(t) i_t\s x\s, i_\mu\s \Sigma\s(t) i_t\s
 x\s]_{H_\mu}
 \\ &  = \lb Q_t x\s, x\s\rb
 + \lb Q_\infty  \Sigma\s(t) i_t\s x\s, \Sigma\s(t) i_t\s x\s\rb
 \\ &  \le \lb Q_t x\s, x\s\rb + \n Q_\infty\n_{\calL(E\s,E)}  \n \Sigma(t)\n_{\calL(E,H_t)}^2 \n i_t\s
  x\s\n_{H_t}^2
 \\ &  = \bigl(1+ \n Q_\infty\n_{\calL(E\s,E)}  \n \Sigma(t)\n_{\calL(E,H_t)}^2 \bigr) \lb Q_t x\s, x\s\rb.
\end{aligned}
$$
The desired inclusion now follows from \eqref{eq:RKHS}.
\end{proof}

Summarizing our discussion we have proved:

\begin{theorem} If the transition semigroup generated by $L$ is strongly Feller, then
$$\sigma(L_p) =
{\Bigl\{ \sum_{j=1}^n k_j z_j: \ \ k_j\in\N, \ z_j
\in\sp(A_\mu);\ j=1,\dots,n; \ n\ge 1\Bigr\}}.
$$
\end{theorem}

\subsection{The finite-dimensional case}\label{sec7}
We will show next that the assumptions on $S_\mu$
in Theorem \ref{thm:sp-OU}
are also satisfied if $\dim E < \infty$.
 Since in finite dimensions every strongly continuous semigroup
is compact, we only have to check the strict contractivity assumption.

In the following result we do not {\em a priori} assume Hypothesis
(H$Q_\infty$).
We identify $E$ and its dual in the natural way.

\begin{proposition}\label{prop:invertible}
Let $\dim\,E <\infty$. The following assertions are
equivalent:
\begin{enumerate}
\item
$Q_\infty:=\lim_{t\to\infty} Q_t$ exists
and 
$Q_\infty$ is invertible;
\item
$Q_t$ is invertible for all $t>0$
and $S$ is uniformly exponentially stable.
\end{enumerate}
In this situation we have $H_t = H_\mu=E$ with equivalent norms
and $S_\mu$ is a strict contraction semigroup.
\end{proposition}

\begin{proof}

(1)$\Rightarrow$(2): \
From
rank$\,Q_\infty=n$ and $\lim_{t\to\infty} Q_t = Q_\infty$ we have $\hbox{rank}\,Q_t=n$ for large enough $t$.
For these $t$ we have $H_t = \hbox{range}\,Q_t = E$.
On the other hand,
since the subspaces $H_t$ increase with $t$
and since their dimensions can make only finitely many jumps,
there is a time $t_0>0$
such that $H_t = H_{t_0}$ for all $0<t\le t_0$.
It then follows from \cite[Theorem 1.4]{Ne} that $S(s)$ maps $H_{t_0}=
\hbox{range}\,Q_{t_0}$ into itself
for all $s\ge 0$. The identity
$$Q_{kt_0} = Q_{t_0} + \dots + S((k-1)t_0)Q_{t_0}S\s((k-1)t_0)$$
then implies that
$$H_{kt_0} = \hbox{range}\,Q_{kt_0}
\subseteq \hbox{range}\,Q_{t_0}= H_{t_0} = H_t$$
for all $k\ge 1$. But by the observations already made,
for $k$ large enough we have $H_{kt_0} = E$ and therefore
$H_t = E$ for all $0<t\le t_0$. But then we have $H_t=E$ for all $t>0$.
This means that $Q_t$ is invertible for all $t>0$.

The above arguments show that for all $t>0$,
$H_t=H_\mu=E$ with equivalent norms.
By Lemma \ref{lem:str-contr}, the first of these identities
implies that $S_\mu$ is a strict contraction
semigroup. The second of these identities then implies that $E$ can be renormed
in such a way that $S$ is a strict contraction group. In particular, $S$ is
uniformly exponentially stable.

(2)$\Rightarrow$(1): \  The existence of the limit defining $Q_\infty$ is
obvious from the uniform exponential stability of $S$. The invertibility of
$Q_\infty$ follows the inclusion
$H_t\subseteq H_\mu$: the invertibility of $Q_t$ means that $H_t=E$ and
therefore $H_\mu=E$.
\end{proof}

It is possible (but not entirely straightforward) to give 
a direct proof of the strict
contractivity of $S_\mu$ based on a compactness argument and some
elementary spectral theory.

In the results so far we could assume without loss of generality that
$\mu$ is nondegenerate, but this was nowhere essential. In our final result,
the nondegeneracy is crucial:

\begin{theorem}\label{thm:fd}
If $\dim E<\infty$ and
$\mu$ is nondegenerate,
then for all $p\in (1,\infty)$ we have
$$
\sp(L_p) =
{\Bigl\{ \sum_{j=1}^n k_j z_j: \ \ k_j\in\N, \ z_j
\in\sp(A);\ j=1,\dots,n; \ n\ge 1\Bigr\}}.
$$
\end{theorem}

\begin{proof} Since $\dim E<\infty$,
the nondegeneracy of $\mu$ implies
the invertibility of $Q_\infty$. Consequently we have $H_\mu=E$ with
equivalent norms, and therefore $\sp(A_\mu)=\sp(A)$. The result now follows from the second part of Theorem \ref{thm:sp-OU}.
\end{proof}
In  \cite{MPP}, Theorem \ref{thm:fd} was proved
under the assumptions that $\dim E < \infty$, $Q_t$ is invertible 
for all $t>0$, and $S$ is
uniformly exponentially stable. 
By Proposition \ref{prop:invertible} and the fact that in finite dimensions every
positive symmetric operator is a Gaussian covariance, 
these assumptions are equivalent to the ones of Theorem
\ref{thm:fd}, viz. the existence of a nondegenerate invariant measure.

It was observed in \cite{MPP} that all generalized eigenvectors of $L_p$
are polynomials. This fact follows effortless
from our approach. First, by Theorem
\ref{thm:sp-OU} and compactness, every $\l\in \sp(L_p)$ is an
eigenvalue of finite multiplicity.
If $f$ is a generalized eigenvector, then for all $n$ sufficiently large,
$\sum_{j=0}^n I_{p,n}f$ is nonzero and therefore
a generalized eigenvector as well. But since the generalized
eigenspace of $\l$
is finite-dimensional, at most finitely many $I_{p,n}f$ are nonzero.
Hence, $f = \sum_{k=1}^m I_{p,n_k}f$ with each $I_{p,n_k}f\in \H_{p,n_k}$.
But each $\H_{p,n_k}$ is the linear span of a finite set of
polynomials, and therefore $f$ is a polynomial.

\end{document}